\documentclass[preprint,12pt]{elsarticle}

\usepackage{amssymb}
\usepackage{subfigure}
\usepackage{graphics}
\graphicspath{ {./images/} }
\usepackage{epsfig}
\usepackage{epstopdf}
\usepackage{amsmath}
\usepackage{array}
\usepackage{graphicx}

\journal{Physics Letters B}

\begin{document}

\begin{frontmatter}



\title{Raychaudhuri Equation in Kaluza-Klein Space-Time}


\author{Vishnu S Namboothiri}
\address{Centre for Science in Society, Cochin University of Science and Technology, Cochin - 682022, India, \newline ramharisindhu@gmail.com}

\begin{abstract}
We studied the Raychaudhuri equation in Kaluza-Klein space-time. We derived an additional term that is solely due to Kaluza-Klein's unification. This term affects the defocus of world lines near the singularity of charged higher-dimensional black holes. We illustrated it with a static higher dimensional black hole generated by Einstein-Maxwell equations. It has been found that the term derived modifies the repulsive effect of kaluza Klein scalar on world lines. We proved some mathematical theorems which give an insight into the interior of the black holes.
\end{abstract}

\end{frontmatter}


\section{Introduction}
\label{section 1}

 Kaluza-Klein theory is a geometrical unification of electromagnetism and gravitation. Kaluza and Klein were the two scientists who were behind this theory. Here the geometric unification is achieved by postulating the principle of general relativity in five-dimensional space-time. We get: $R^{5}=R^{4}-\frac{1}{4}F^{\mu\nu}F_{\mu \nu}$.
Where, $R^{5}$ Ricci scalar in five dimension, $R^{4}$ is Ricci scalar in four dimension and  $\frac{1}{4}F^{\mu\nu}F_{\mu \nu}$ as electromagnetic Lagrangian. This equation is known as the Kaluza Klein miracle. 

Even though the Kaluza-Klein theory is an ambitious task in unifying electromagnetism with gravity, it has some drawbacks. The major drawback is the compactification of extra dimensions. The compactification of extra dimension implies a tower of particles with increasing the principal quantum number controlled by compactification length. 
This process will take place in the order of the energy of Tev. So it has not been observed yet. Kaluza Klein theory is a precursor of String theory.

The presence of the electromagnetic field in the right-hand side of  Einstein's field equations that matters part will affect the Ricci scalar and then it has an effect on the dynamics of world lines. This is obvious in standard General relativity. But this effect has not much studied near singularities using the Raychaudhuri equation\cite{Raychaudhuri:2000pd}\cite{Brout:1977ix}. Raychaudhuri equation has been important in dealing with singularities\cite{Hawking:1973uf}. Papers of Witten\cite{Witten:2019qhl},Geroch \cite{Geroch:1968ut}, book written by padmanabhan \cite{Padmanabhan:2010zzb},Sayankar\cite{Kar:2008zz} and \cite{Koley:2003tp}. Further the classic works of Hawking ,Penrose and Ellis gives mathematical theorems on singularities. Further works of Dadhich \cite{Dadhich:2007pi}  \cite{Dadhich:2005qr} gives account of current status of Raychaudhuri equation.

\par Since it gives a complete picture of dynamics near the singularity. Here we study a cosmology coupled with a scalar in the context of Klauza-Klein theory. This scalar effects to defocus the world lines near the singularity. This has been studied by R Parthasarathy.Investigation of Kaluza-Klein theory leads to a vast literature. In this note, we review the paper "Classical defocusing of world lines in higher dimensions" by R.   Parthasarathy\cite{Parthasarathy:2017nye}.  This paper discusses the positive contribution of Kaluza-Klein scalar in the Raychaudhuri equation and thereby causing defocusing of world lines near singularities.  It has an application in cosmology in such cases where the universe has an infinite age hence no beginning and thereby avoiding the singularity. In the case of the Raychaudhuri equation, there is a lot of review articles\cite{Dadhich:2005qr}.  Quantum corrected models were studied by S Das\cite{Das:2013oda}.

 \section{Effect of Electromagnetism in Raychaudhuri equation }
 \label{section 2}
In this paper, we are investigating the classical properties of the Kaluza-Klein theory and discuss the charged black holes. The additional term derived from Kaluza Klein theory can be applied to black holes generated by Einstein-Maxwell equations as well as other five-dimensional charged black holes(eg: Five-dimensional Reissner–Nordström metric) so on.
Singularities are one of the problems in classical gravity.   There are quantum corrected models in avoiding these singularities\cite{Das:2013oda}.  In these models, this is achieved through quantum corrections to the Raychaudhuri equation by replacing classical geodesics with Bohmian trajectories and thus prevent focusing of geodesics. Within the classical theory itself, there is an alternative mechanism for avoiding these singularities.  We can solve the problem of singularities within the framework of classical theory if we extend the standard general relativity to five dimensions\cite{Parthasarathy:2017nye}.

In the case of 4-dimensional formulation, we introduce electromagnetism by hand in the geodesic equation. 
 Thus the geodesic equation  in the presence of electromagnetic fields  would be the Lorentz force law in the covariant form: 
 \begin{equation}
\frac{d^2z^\mu}{dS^2}+{\Gamma}^\mu_{\nu \rho}\ \frac{dz^\nu}{dS}\ \frac{dz^\rho}{dS}=(q/mc)F^{\mu\nu}u_{\nu},
\end{equation}
with
$$u_{\nu}=\frac{dx^{\nu}}{ds}.$$
Consider the following Kaluza-Klein metric, $g_{AB}, A,B=0,1,2,3,5 $ with the  electromagnetic potential
\begin{equation}
g_{AB}=
  \left[ {\begin{array}{cc}
 g_{\mu \nu}+\alpha^{2}\:g_{55}A_{\mu}A_{\nu} &\alpha g_{55}A_{\mu}\\ 
  \alpha g_{55}A_{\nu}  & g_{55} \\
  \end{array} } \right]
\end{equation}
where $g_{\mu \nu}$ is the 4-dimensional metric. Now the line element becomes:
\begin{equation}
 dS^2=g_{\mu\nu}\ dx^{\mu}dx^{\nu}-g_{55}\ (dx^5+ \alpha A_{\mu}dx^{\mu})^2.  
\end{equation}
We also have 
\[ 
g^{AB}=
  \left[ {\begin{array}{cc}
 g^{\mu \nu} &-\alpha g^{\mu\nu}A_{\nu}  \\
  -\alpha g^{\nu\mu}A_{\mu}  & 1/g_{55}+\alpha^{2}\:g^{\mu\nu}A_{\mu}A_{\nu} \\
  \end{array} } \right].\]
This provides a space-time with electromagnetism. The  geodesic equation in five-dimensional space-time,  
\begin{equation}
\frac{d^2z^A}{dS^2}+{\Gamma}^A_{B C}\frac{dz^B}{dS}\frac{dz^C}{dS}=0  
\end{equation}
 can be transformed by applying cylindrical condition on the metric as:
\begin{equation}
\frac{d^2z^{\mu}}{dS^2}+{\Gamma}^{\mu}_{\nu\lambda}\ \frac{dz^{\nu}}{dS}\ \frac{dz^{\lambda}}{dS}=a\alpha F_{\mu\nu}\frac{dz^{\nu}}{dS}\\
+\frac{1}{2} 
\frac{a^2}{g_{55}^2}g^{\mu\lambda}\left({\partial}_{\lambda}g_{55}\right), \end{equation}
where
\begin{equation}
a=g_{5\mu}\frac{dz^\mu}{dS}+g_{55}\frac{dz^5}{dS}, \nonumber 
\end{equation}
and $a$ is a constant along the $5-D$  world line.  In our case $g_{5\mu}$  is not zero since we have included electromagnetic fields.  We assume that:
$$ \frac{g_{5\mu}}{g_{55}}=\alpha A_{\mu}(x) $$
where $\alpha$ is determined by
$$ \alpha= \frac{q}{amc}.$$
Here the electromagnetic potential emerges out of $g_{5\mu}$.  Let us now consider Raychaudhuri equation in four dimensions
\begin{equation}
\frac{d\Theta}{dS}=-\frac{{\Theta}^2}{3}-{\sigma}_{\mu\nu}{\sigma}^{\mu\nu}+{\omega}_{\mu\nu}{\omega}^{\mu\nu}
-R_{\mu\nu}u^{\mu}u^{\nu}+(\dot{u}^{\mu})_{;\mu}
\end{equation}
 where $\displaystyle u^{\mu}=\frac{dz^{\mu}}{dS}$, $2{\sigma}^2={\sigma}_{\mu\nu}{\sigma}^{\mu\nu}$, $2{\omega}^2={\omega}_{\mu\nu}
{\omega}^{\mu\nu}$.  The cosmological constant $\Lambda$ is set to zero. 
\cite{Parthasarathy:2017nye}
Since, $$\dot{u^{\mu}}=\frac{d^2z^{\mu}}{dS^2}+\Gamma^{\mu}_{\alpha \beta}u^{\alpha}u^{\beta},$$  the last term in Eq.(7)can be written as: 
\begin{equation}
(\dot{u}^{\mu})_{\mu}=-\frac{a^2}{2}\ g^{\mu\rho} D_{\mu}({\partial}_{\rho}\frac{1}{g_{55}})+a \alpha D_{\mu}(F_{\mu\nu}\frac{dz^{\nu}}{dS}) 
\end{equation}
with  $D_{\mu}$ as the covariant derivative. 
The vorticities $\omega$  and $\sigma$ induces expansion and contraction respectively. \cite{Parthasarathy:2017nye}. It is useful to note that $-D_{\mu}({\partial}_{\rho}\frac{1}{g_{55}})$ is positive for static spherically symmetric space-time in five dimensions without electromagnetism\cite{Parthasarathy:2017nye}.  Now 
additional term, $-D_{\mu}({\partial}_{\rho}\frac{1}{g_{55}})$ is positive for static spherically symmetric space-time in five dimensions without electromagnetism\cite{Parthasarathy:2017nye}.  Now 
consider a case with $R_{{\mu}{\nu}}u^{\mu}u^{\nu}>0$ and $\omega=0$
we get that. 
\begin{equation}
\frac{d\Theta}{dS}\leq\frac{1}{3}\Theta^2-\frac{a^2}{2}\ g^{\mu\rho}D_{\mu}({\partial}_{\rho}\frac{1}{g_{55}})+a \alpha D_{\mu}(F_{\mu\nu}\frac{dz^{\nu}}{dS})
\end{equation}

\section{\label{sec:level4} Kaluza-Klein black hole}

We are now giving a brief review of the work done by  H. Ishihara and K.Matsuno. Most of the results taken from this paper and that have importance in investigating defocusing or focusing of world lines near the singularity
We now examine the Einstein-Maxwell theory described 
by the action
\begin{equation}
S=\frac{1}{16\pi G}\int d^4x \sqrt{-g}\left( R 
    -F_{\mu\nu}F^{\mu\nu} \right),
\label{action}
\end{equation}.
Where $R$ is the scalar curvature, $G$ denotes the Newton's constant and $A_{\mu}(r,t)$ is the gauge potential. From the  equations of motion,
\begin{equation}
R_{\mu\nu}- \frac12R g_{\mu\nu}
    = 2\left(
        F_{\mu\lambda}F_\nu^{~\lambda}
        -\frac14 g_{\mu\nu}F_{\alpha\beta}F^{\alpha\beta}\right). 
\end{equation}
and its solution 
\begin{equation}
dS^2 = -f dt^2 +  \frac{k^2}{f} dr^2
       +\frac{r^2}{4}\left[ 
     k \left\{ (\sigma^1)^2+(\sigma^2)^2 \right\} + (\sigma^3)^2 \right],
\end{equation}
where $f$ and $k$ are functions of $r$ defined by
\begin{equation}
f(r)= \frac{(r^2-r_+^2)(r^2-r_-^2)}{r^4}\:,
k(r) = \frac{(r_{\infty}^2-r_{+}^2)(r_{\infty}^2-r_{-}^2)}{(r_{\infty}^2-r^2)^2}
\end{equation}
and the gauge potential is

\begin{equation}
A=\pm\frac{\sqrt{3}}{2}\frac{r_+r_-}{r^2} dt. 
\end{equation}
Here, $r_{\pm}$ and $r_{\infty}$ are constants, and these $r_{+}$ and $r_{-}$ can be interpreted as outer and inner horizon respectively
and $\sigma^i~ (i=1,2,3)$ satisfy the equation.
\begin{equation}
     d\sigma^{i}=\frac{1}{2}C^i_{jk}\sigma^{j}\wedge \sigma^{k}.
\end{equation}
With $ C^{1}_{23}=C^{2}_{31}=C^{3}_{12}=1$ and $C^i_{jk}=0$, for all other cases. The static space-time of the metric has isometry group $SO(3) \times U(1)$.\cite{Ishihara:2005dp}.
\subsection{Properties of Kaluza Klein charged Black hole}

The feature of this black hole is its geometrical structure. These black holes have horizons in the form of squashed  $S^{3}$. Their asymptotic structure consists of twisted $S^{1}$ over the flat space-time at the spatial infinity.
The space-time is five-dimensional in the vicinity of the black hole and four-dimensional with a compact extra dimension at infinity.

\par
Let us consider the shape of horizons of the black hole:
A time $t= constant$  of the space-time that is orthogonal to the time -like killing vector, is decorated by the three-dimensional surfaces, which are specified by $r=constant$; say $\Sigma_{r}$. Each surface $\Sigma_{r}$ is defined as a $S^{1}$  over $S^{2}$ base space.
The surface $\Sigma_{r}$ takes the form on which $ SO(3) \times U(1)$ act as an isometry group. The metric is

\begin{equation}
    ds^2_{\Sigma_r}=\frac{r^2}{4}\left[ 
     k(r) \left\{ (\sigma^1)^2+(\sigma^2)^2 \right\} +(\sigma^3)^2 \right] 
     =    \frac{r^2}{4}\left[ k(r) d\Omega_{S^2}^2 + \chi^2 \right] ,
\label{Sigma_metric}
\end{equation}
where
\begin{equation}
d\Omega_{S^2}^2=d\theta^2+\sin^2\theta d\phi^2, \quad
    \chi=\sigma^3=d\psi+\cos\theta d\phi. \notag \\                  
    (0 \leq \theta < \pi, ~0 \leq \phi < 2\pi,~ 0 \leq \psi < 4\pi) \label{base_fiber}
\end{equation}
The main important point is the aspect ratio of the $S^{2}$ base space to the $S^{1}$ fiber is denoted by $k(r)$.
This $k(r)$ denotes the squashing of $S^{3}$. The degree of squashing increases monotonically as $r$ increases towards $r_{\infty}$.
At $r_{\infty}$ the function($k(r)$) diverges and this can be interpreted as collapse of $S^{3}$ to $S^{2}$ there. The shape of time-slices of the outer and inner horizons is described by three-dimensional metrics, with $r=r_{\pm}$ respectively.The definition of the squashing function is as follows
\begin{equation}
     k(r_{\pm}) =  \frac{r_{\infty}^2-r_{\mp}^2}{r_{\infty}^2-r_{\pm}^2}.
\end{equation}.
From this, we can interfere that the outer horizon is 'oblate' with $S^{2}$ larger than $S^{1}$ and the inner horizon is 'prolate' with  $S^{2}$ smaller than $S^{1}$. In the degenerate case, $r_{+}=r_{-}$, the shape of the horizon is the round in $S^{3}$.\cite{Ishihara:2005dp}

\subsection{Asymptotic structure and Physical properties of Kaluza-Klein Black hole}
\subsubsection{Asymptotic structure}
There seem two singularities in the black hole. One is at $r=0$ and at $r=r_{\infty}$. But the singularity at $r=r_{\infty}$ is spatial infinity. We have to change from the old coordinate system to the new one. To observe this, put
\begin{equation}
\rho =\rho_{0} \frac{r^2}{r_{\infty}^2-r^2}, 
\end{equation}
Where
\begin{equation}
\rho_{0}^2 = k_{0} \frac{r_{\infty}^2}{4}=
    k_{0}=k(0)= f_{\infty}= f(r_{\infty})
\end{equation}
But,
\begin{equation}
f(r_{\infty})= \frac{(r_{\infty}^2-r_{+}^2)(r_{\infty}^2-r_{-}^2)}{r_{\infty}^4} 
\end{equation} 
The new introduced coordinate $\rho$ varies from $0$ to $\infty$ 
when $r$ varies from $0$ to $r_{\infty}$. The metric can be written as 
\begin{equation}
    ds^2 = - V dT^2 +\frac{K^2}{V} d\rho^2
        +R^2 d\Omega_{S^2}^2 + W^2 \chi^2, 
\end{equation}

where $V, K, R$ and $W$ are functions of $\rho$ in the form
\begin{equation}
V=\frac{(\rho-\rho_+)(\rho-\rho_-)}{\rho^2}, \quad
    K^2 = \frac{\rho+\rho_{0}}{\rho}, \\
    R^2=\rho^2 K^2 ,\quad
    W^2=\frac{r_{\infty}^2}{4}~K^{-2}
        =(\rho_{0}+\rho_{+})(\rho_{0}+\rho_{-})~ K^{-2}.
\end{equation}
Now the metric is in new coordinates $\rho$ and $T=\sqrt{f_{\infty}}t$.
From the new parameters we can write
\begin{equation}
\rho_\pm = \rho_{0} \frac{r_{\pm}^2}{r_{\infty}^2-r_{\pm}^2}.
\end{equation}
If $\rho\rightarrow\infty$, i.e., $r \rightarrow r_{\infty}$, 
the metric 21 with 22 approaches 
\begin{equation}
    ds^2 =-dT^2 + d\rho^2
        +\rho^2 d\Omega_{S^2}^2 + \frac{r_{\infty}^2}{4} \chi^2.
\end{equation}
In the limit of large $\rho$, the leading order term of the  Kretschmann scalar is
\begin{equation}
    R_{\mu\nu\lambda\sigma}R^{\mu\nu\lambda\sigma} 
        \sim \frac{12 \left[ (\rho_{+}+\rho_{-})^2 -\rho_{+}\rho_{-} 
        +2(\rho_{0}+\rho_{+})(\rho_{0}+\rho_{-}) \right] }{\rho^6} .
\end{equation}

In the limit, $\rho \rightarrow \infty$ or
$ r \rightarrow r_{\infty}$, corresponding to spatial infinity, and space-time is locally flat. It can be put in this way topologically not a direct product but a twisted $Sp1$ fiber bundle
over four-dimensional Minkowski spacetime.\cite{Ishihara:2005dp}.
\subsubsection{Physical properties}
From metric 12, we can say that in the near region $ r \leq r_{i} $ the function $ k(r) $ is the nearly constant and therefore hyper area is proportional to $ r^3 $. For $ r \leq r_{i} $ space-time would be similar to Reissner-Nordstrom black hole. For the case $ r_{+} \leq r_{i} $, there exists an outer region 
 $r_{+} \leq r \leq r_{\infty}$, then
the space-time would behave like that of a five-dimensional black hole for observers in this space-time.
Inside the horizon $\Sigma_{r}$ shrinks to a point with the constant $k_{0}$ as $r\rightarrow 0$. 
Surely, a time-like singularity exists at $r=0$,  
as in the case of the Reissner-Nordstr\"om metric. Now the Kretschmann scalar diverges in the limit $r \rightarrow 0$. 
\begin{equation}
  R_{\mu\nu\lambda\sigma}R^{\mu\nu\lambda\sigma} 
        \sim 508 r_+^4r_-^4/\left( k_0^8r^{12} \right)   
\end{equation}
The value $k_{0}$ denotes the elongation of central singularity.
\par More Physical properties of the above discussed black hole can be found in H. Ishihara and K.Matsuno, "Kaluza-Klein black holes with squashed horizons" in the bibliography\cite{Ishihara:2005dp}.

\section{Analysis with Kaluza-Klein Black hole}
Here we a given theorem, it's proof and Physical significance.
\subsection{Theorem}

\textbf{  In the  metric of the form $ dS^2 = -f dt^2+ \frac{k^2}{f} dr^2 +\frac{r^2}{4}\left[  k \left\{ (\sigma^1)^2+(\sigma^2)^2 \right\} + (\sigma^3)^2 \right]$ the defocusing or focusing effects due to $ +a \alpha D_{\mu}(F_{\mu\nu}\frac{dx^{\nu}}{dS})$ occurs only when $r\geq \sqrt{\frac{r_{+}^2r_{-}^2}{r_{+}^2+r_{-}^2}}$  under radial congruence and marginally bounded case of geodesics. }

\setlength{\parskip}{1em}
 \textbf{Proof}
 According to equation 13, the vector potential has only $A_{t}$ component is nonzero. For this gauge only electric field component exists like $F_{tr},F_{t\theta},F_{t\phi}$. Other components like $F_{r\theta}$ are zero. Moreover $F_{\mu\nu}$ is anti-symmetric. But $F_{rt}= \pm\frac{\sqrt{3}r_{+}r_{-}}{r^3},F_{t\theta}=0,F_{t\phi}=0$. For radial congruence  $\frac{dz^{\theta}}{dS}=0,\frac{dz^{\phi}}{dS}=0$. Geodesics are marginally bound means the the rest mass energy is conserved energy. $E=-u_{\alpha}\xi^{\alpha}_{t}=-u_{t}=1$, this implies $u^{t}=\frac{1}{f}$. From normalization condition:
 $g_{\alpha\beta}u^{\alpha}u^{\beta}=-1$
 we get $u^{r}=\sqrt{1-f}$ where $f(r) = \frac{(r^2-r_{+}^2)(r^2-r_{-}^2)}{r^4}$.  So $\sqrt{1-f(r)}=\sqrt{(\frac{r^2(r_{+}^2+r_{-}^2)-r_{+}^2r_{-}^2}{r^4})}$. This implies that shows $1-f\geq 0$($u^{r}$ must be positive and real) and $r\geq \sqrt{\frac{r_{+}^2r_{-}^2}{r_{+}^2+r_{-}^2}}$. 
 \par
 We have:
 \begin{equation}
 F_{\mu}=F_{\mu\nu}\frac{dz^{\nu}}{dS}
 \end{equation}
 We can use the formula $D_{\mu}F_{\mu}=\partial_{\mu}F_{\mu}-\Gamma^{\lambda}_{\mu\mu}F_{\lambda}$.\\
After some algebraic steps we get an important result, $a \alpha D_{\mu}(F_{\mu\nu}\frac{dz^{\nu}}{dS})$ as
\begin{equation}
 \pm a\alpha \sqrt{3}r_{+}r_{-}(\partial_r(\frac{1}{fr^3})-\frac{1}{2k^2r^3}\partial_{r}(\frac{k^2}{f}))
\end{equation}

\subsection{Physical significance}
We have already discussed the physical properties of the black hole in the previous section. Here we have to discuss the physical significance of the above theorem and result.  When the distance becomes $r\geq \sqrt{\frac{r_{+}^2r_{-}^2}{r_{+}^2+r_{-}^2}}$, then the radial velocity becomes zero. After that radial velocity becomes imaginary. So we have to say that under radial congruence it is the limiting distance and before that geodesics should focus or defocus away from the singularity.
\par 
 Here it is due to the property of Kaluza-Klein space-time. Of course, the electromagnetic field of some charged black holes in four dimensions shows an effect in the motion of particles around it. \cite{Sharif:2017twj}.  The geometrical way of explaining to defocus or focusing on world lines due to charge distribution near-singularity is not much studied.
 \par 
 This term ($+a \alpha D_{\mu}(F_{\mu\nu}\frac{dx^{\nu}}{dS})$) introduces the fact that the electric field can affect the defocusing and focusing of geodesics depending on the sign in front of gauge field $A_{t}(r)$. So the focus and defocus also depend on the nature of the electromagnetic potential.
 \par
 consider the Kaluza-Klein scalar $\psi(r)=\frac{r}{2}$. So according to affect of Kaluza Klein scalar($-\frac{a^2}{2}\ g^{\mu\rho} D_{\mu}({\partial}_{\rho}\frac{1}{g_{55}})$ Here $g_{55}=\frac{r}{2}$) will be positive and as a result world lines always defocus near singularity\cite{Parthasarathy:2017nye}.The electromagnetism will modify this defocus of world lines. But In this particular black hole defocus or focus depends on several factors like $r_{+}$,$r_{-}$,$r_{\infty}$ and charge of coming particle\:($\alpha$).

\section{Brief Discussion of results}

Anyway, the results we proved are true for static higher-dimensional black holes generated by Einstein-Maxwell equations.
We give further points by taking different limits of $r_{+},r_{-},r_{\infty}$.
\par
First, we take the limit $ r_{\infty} \rightarrow \infty $. The squashing function behaves as $k \rightarrow  1$ in this limit, and
therefore the metric reduces to that of the five-dimensional ReissnerNordstrom black hole.
In this case,The effect due to electromagnetism on world lines occurs.
\par
Second, in the limit, $ r_{-} \rightarrow 0$, the metric describes a neutral black hole with a gravitational radius $r_{+}$. In that case  $a \alpha D_{\mu}(F_{\mu\nu}\frac{dz^{\nu}}{dS})$   vanishes.
\par
Third, when we set $r_{-} = r_{+}$, the metric describes a deformed   five-dimensional
extremal Reissner-Nordstrom black hole with a degenerate horizon($ r_{-} = r_{+} $). In that case,
The above terms effects and the theorem we proved can be applied:
\begin{equation}
  r \geq \frac{r_{+}}{\sqrt{2}} 
\end{equation}
Thus the defocus of world lines occurs in this inequality.
 The metric obtained by suitable coordinate transformation can be studied as super-symmetric solution of five dimensional super gravity\cite{Ishihara:2005dp}. 
 \par
Next, we consider the limit $ r_{+},r_{-} \rightarrow r_{\infty} $, with $\rho_{\pm}$ finite. In this case This metric describes a four-dimensional ReissnerNordstrom black hole with a twisted S1 bundle,
where the size of the S1 fiber takes the constant value $\sqrt{\rho_{+}\rho_{-}}=\frac{r_{\infty}}{2}$ \cite{Ishihara:2005dp}.
In this case, also, We can say that the above inequality, as well as deflection term, can be applied. This is true as $r_{+}$ and $r_{-}$ is
not zero or in other words when a black hole is charged.

\section{Conclusion}
The formulae ($+a \alpha D_{\mu}(F_{\mu\nu}\frac{dx^{\nu}}{dS})$) we derived in this work can be extended to the charged black hole in higher dimensions or black holes formed with dimensional reduction.
From the above discussion, this term ($+a \alpha D_{\mu}(F_{\mu\nu}\frac{dx^{\nu}}{dS})$) as well as the theorem we proved in radial congruence, marginally bounded case gives us an insight about the geometry of singularity in black holes.
\par
Future works can be done in this regard is the study of magnetic monopole and extending these results into rotating black holes in higher dimensions.

\section{Acknowledgement}
I am very much thankful to Prof R. Parthasarathy (CMI), Prof A.P. Balachandhran (Syracuse University), Karthik Rajeev (IUCAA), Prof Naresh Dadhcih (IUCAA) for comments and discussions on this work. I always remember the help and motivation of Prof Ramesh Babu T. (CUSAT), Prof V.P.N. Nampoothiri(CUSAT). This work is dedicated to my parents and friends.





\end{document}